\documentclass[aps,prl,reprint,twocolumn,superscriptaddress,floatfix,nofootinbib,longbibliography]{revtex4-1}
\usepackage{graphicx,amsmath,amsfonts,amssymb,amsthm,xr}
\usepackage{epsfig,amsmath,amssymb,color,dsfont,upgreek,physics}
\usepackage{mathrsfs}
\usepackage{mathtools}
\usepackage{bbold}
\usepackage{comment}
\usepackage{float}

\usepackage[bookmarks=true,colorlinks,linkcolor=OrangeRed,urlcolor=NavyBlue,citecolor=RoyalBlue]{hyperref}
\usepackage[dvipsnames]{xcolor}
\usepackage{orcidlink}

\definecolor{mygold}{rgb}{0.93,0.69,0.13}

\definecolor{mypurple}{rgb}{0.49,0.18,0.56}

\definecolor{mygreen}{rgb}{0,0.5,0}

\definecolor{mygreen}{rgb}{0,0.5,0}
\definecolor{myred}{rgb}{0.7,0,0}
\definecolor{myblue}{rgb}{0,0,1}

\begin{document}
\title{Probing Confinement Through Dynamical Quantum Phase Transitions:\\From Quantum Spin Models to Lattice Gauge Theories}
\author{Jesse Osborne${}^{\orcidlink{0000-0003-0415-0690}}$}
\email{j.osborne@uqconnect.edu.au}
\affiliation{School of Mathematics and Physics, The University of Queensland, St.~Lucia, QLD 4072, Australia}
\author{Ian P.~McCulloch${}^{\orcidlink{0000-0002-8983-6327}}$}
\email{ian@phys.nthu.edu.tw}
\affiliation{Department of Physics, National Tsing Hua University, Hsinchu 30013, Taiwan}
\affiliation{School of Mathematics and Physics, The University of Queensland, St.~Lucia, QLD 4072, Australia}
\author{Jad C.~Halimeh${}^{\orcidlink{0000-0002-0659-7990}}$}
\email{jad.halimeh@physik.lmu.de}
\affiliation{Department of Physics and Arnold Sommerfeld Center for Theoretical Physics (ASC), Ludwig-Maximilians-Universit\"at M\"unchen, Theresienstra\ss e 37, D-80333 M\"unchen, Germany}
\affiliation{Munich Center for Quantum Science and Technology (MCQST), Schellingstra\ss e 4, D-80799 M\"unchen, Germany}

\begin{abstract}
Confinement is an intriguing phenomenon prevalent in condensed matter and high-energy physics. Exploring its effect on the far-from-equilibrium criticality of quantum many-body systems is of great interest both from a fundamental and technological point of view. Here, we employ large-scale uniform matrix product state calculations to show that a qualitative change in the type of dynamical quantum phase transitions (DQPTs) accompanies the confinement--deconfinement transition in three paradigmatic models --- the power-law interacting quantum Ising chain, the two-dimensional quantum Ising model, and the spin-$S$ $\mathrm{U}(1)$ quantum link model. By tuning a confining parameter in these models, it is found that \textit{branch} (\textit{manifold}) DQPTs arise as a signature of (de)confinement. Whereas manifold DQPTs are associated with a sign change of the order parameter, their branch counterparts are not, but rather occur even when the order parameter exhibits considerably constrained dynamics. Our conclusions can be tested in modern quantum-simulation platforms, such as ion-trap setups and cold-atom experiments of gauge theories.
\end{abstract}

\date{\today} 
\maketitle

\textbf{\textit{Introduction.---}}Confinement is a prominent effect in physics, most famous in the context of color confinement in quantum chromodynamics (QCD). Below the Hagedorn temperature, color-charged particles, such as quarks and gluons, cannot be isolated and are confined into hadrons \cite{Ellis_book}. Confinement is also present in the context of another gauge theory, quantum electrodynamics (QED). There, a topological $\theta$-angle \cite{tHooft1976,Jackiw1976,Callan1979} or the gauge coupling \cite{Chandrasekharan1999} can be tuned to induce confinement. Recently, there has been significant interest in probing confinement of gauge theories on dedicated large-scale quantum simulators \cite{Zohar2012,Mildenberger2022,Halimeh2022tuning,Cheng2022tunable,Zhang2023observation}, including at finite temperature \cite{Davoudi2022quantum,Fromm2023simulating,Kebric2023confinement}.

Beyond gauge theories, confinement can also arise in the context of quantum spin chains. If one considers a quantum Ising chain with transverse and longitudinal fields, then real-time confinement in the wake of a quantum quench arises, where domain-wall excitations due to the transverse field tend to bind together due to the longitudinal field \cite{Kormos2017}. Confinement also occurs in transverse-field Ising chains with long-range interactions \cite{Liu2019,Halimeh2020quasiparticle,Ranabhat2023dynamical}, where the latter also makes it energetically unfavorable for domain-wall excitations to be separated at long distances. The two-dimensional Ising model also exhibits confinement-like behavior \cite{Belavin1984infinite,Henkel1989}. In two dimensions and with nearest-neighbor interactions, the energy of a domain (or island) of opposite spin polarization in the ferromagnetic phase is proportional to its perimeter, unlike in the one-dimensional case where the cost is independent of domain size. Dynamically, this leads to a suppression of the spread of such domains when the transverse-field strength is not large enough.

Whether in quantum spin models or gauge theories, the presence of confinement gives rise to exotic constrained dynamics that is very interesting to understand from a fundamental point of view, and also to possibly probe on modern quantum simulators \cite{Sedgewick2002,Pichler2016,Neyenhuis2017,Barros2019string,Magnifico2020realtimedynamics,Lerose2019quasilocalized,Tan2021,Vovrosh2021,Lagnese2021}.

In recent years, there has been a concerted effort to build a complete theoretical framework of far-from-equilibrium quantum many-body criticality. Various approaches have been explored, with one particular avenue that has garnered significant attention: \textit{dynamical quantum phase transitions} (DQPTs) \cite{Silva2008,Heyl2013,Heyl_review}. It entails extending the concept of thermal free energy in equilibrium to Schr\"odinger time evolution. By recognizing that the overlap of the time-evolved wave function with the initial state is a boundary partition function, one can interpret evolution time as complexified inverse temperature. The logarithm of this overlap is then related to a \textit{dynamical} free energy whose nonanalyticities at \textit{critical evolution times} are DQPTs. DQPTs have been extensively studied over the last decade in nonintegrable short-range quantum spin chains \cite{Karrasch2013dynamical,Vajna2014disentangling,Andraschko2014dynamical}, long-range models \cite{Zunkovic2016,Halimeh2017dynamical,Zauner-Stauber2017probing,Homrighausen2017anomalous,Zunkovic2018dynamical,Defenu2019dynamical,Uhrich2020out,Halimeh2021dynamical,Corps2022dynamical,Corps2023theory,Deutsch2023macrostates}, topological systems \cite{Vajna2015topological,Schmitt2015dynamical,Sedlmayr2018bulk,Hagymasi2019dynamical,Maslowski2020quasiperiodic,Porta2020topological,Okugawa2021mirror,Sedlmayr2022dynamical,Maslowski2023dynamical}, in higher spatial dimensions \cite{Schmitt2015dynamical,Bhattacharya2017emergent,Weidinger2017dynamical,Heyl2018detecting,DeNicola2019stochastic,Hashizume2020hybrid,Hashizume2022}, at finite temperature \cite{Abeling2016quantum,Bhattacharya2017mixed,Heyl2017dynamical,Sedlmayr2018fate,Lang2018concurrence,Lang2018geometric}, and in gauge theories \cite{Zache2019,Huang2019dynamical,Pedersen2021,Jensen2022,VanDamme2022dynamical,Mueller2023quantum,Pomarico2023dynamical}, and they have been observed experimentally \cite{Jurcevic2017,Flaeschner2018,Nie2020experimental}.

In this study, we ask: Can DQPTs be leveraged as a probe of confinement in quantum many-body models? Given that DQPTs offer a window into out-of-equilibrium criticality, and considering that confinement fundamentally changes the nature of the dynamics of a quantum many-body system, it would be prudent to investigate this question, which we do using state-of-the-art numerical methods. We find that the presence of confinement fundamentally changes the type of DQPTs that dominate the return rate of a quenched quantum many-body system. To demonstrate our findings, we consider three paradigmatic systems: the long-range quantum Ising chain, the two-dimensional quantum Ising model on a square lattice, and spin-$S$ $\mathrm{U}(1)$ quantum link model (QLM) formulations of lattice QED.

\textbf{\textit{Long-range quantum Ising chain.---}}We consider the nonintegrable long-range transverse-field Ising chain (LR-TFIC) with power-law interactions, described by the Hamiltonian
\begin{align}\label{eq:tfim}
    \hat{H}_\text{LR-TFIC} = -J \sum_{j>i} \frac{1}{|j-i|^\alpha} \hat{\sigma}^z_i \hat{\sigma}^z_j - h \sum_i \hat{\sigma}^x_i,
\end{align}
where \(\hat{\sigma}^{x,z}_i\) are the \(x\) and \(z\) Pauli matrices acting on site \(i\), \(J\) is the spin coupling constant, \(h\) is the magnetic-field strength, and \(\alpha\) is the exponent of the power law. The LR-TFIC at ferromagnetic coupling ($J>0$) has a thermal phase transition for $\alpha<2$, a phase transition of the Berezinskii--Kosterlitz--Thouless (BKT) \cite{Thouless1969,Berezinskii1971,Kosterlitz1976} type at finite temperature for $\alpha=2$, and no thermal phase transition for $\alpha>2$ \cite{Dutta2001}. It hosts a quantum phase transition at an $\alpha$-dependent critical value of the transverse-field strength $h$ for any $\alpha$, and this transition is related to the spontaneous breaking of a global $\mathbb{Z}_2$ symmetry \cite{Sachdev_book}. For $\alpha\geq3$, it falls in the short-range universality class. In this work, we consider ferromagnetic coupling (\(J > 0\)) and inverse-square interactions (\(\alpha = 2\)), although we have checked that our forthcoming conclusions are valid for other values of $\alpha$.

We initialize the system in the ground-state manifold at \(h = h_\text{i} = 0\), spanned by the basis generated by \(\ket{\psi^+} = \ket{\cdots\uparrow\uparrow\uparrow\uparrow\uparrow\cdots}\) and \(\ket{\psi^-} = \ket{\cdots\downarrow\downarrow\downarrow\downarrow\downarrow\cdots}\).
We choose the initial state to be \(\ket{\psi(t=0)} = \ket{\psi^+}\) and suddenly quench the magnetic field to \(h = h_\text{f} > 0\) to obtain the time-evolved wave function $\ket{\psi(t)}=\mathrm{e}^{-\mathrm{i}\hat{H}_\text{LR-TFIC}t}\ket{\psi^+}$ at evolution time $t$, where we have set Planck's reduced constant to $\hbar=1$.
To numerically simulate the time evolution of this state, we use infinite matrix product state (iMPS) techniques, which work directly in the thermodynamic limit by writing the state in a translation-invariant form.~\cite{McCulloch:2008,Uli_review,Paeckel_review,mptoolkit}
Specifically, we use a version of the time-dependent variational principle (TDVP) algorithm~\cite{haegeman2011,haegeman2016} that has been modified to handle translation-invariant states.
We use single-site evolution with adaptive bond expansion, with a time-step of \(0.001/J\).
As the inverse-square interaction cannot be directly represented as a matrix product operator (MPO) with finite bond dimension, we instead fit a weighted sum of five exponential functions~\cite{Crosswhite:2008}, which has an efficient representation as an MPO and a total error of less than \(10^{-8}\) over the first $100$~sites.

\begin{figure}[t]
    \centering
    \includegraphics{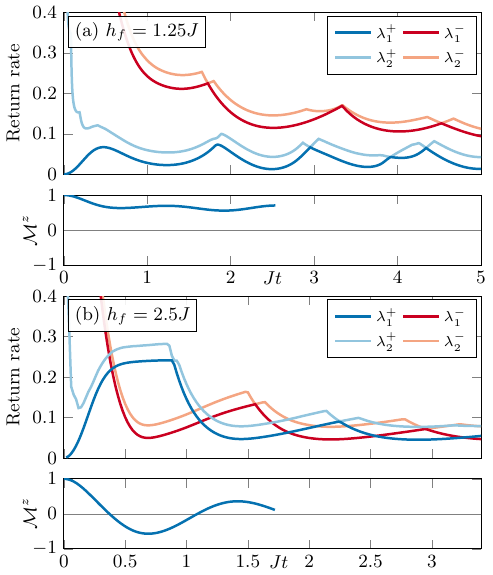}
    \caption{The return rates and order parameter of the quench of the inverse-square long-range LR-TFIC~\eqref{eq:tfim} from \(h_\text{i} = 0\) to (a)~\(h_\text{f} = 1.25J < h_\text{c}^\text{d}\) and (b)~\(h_\text{f} = 2.5J > h_\text{c}^\text{d}\).
    Note that the values of the return rate go to twice the time as those of the order parameter: this is since we can calculate the return rate at time \(2t\) using the wave function at time \(t\) using the doubling trick (see text).}
    \label{fig:tfim}
\end{figure}

To detect the DQPTs in the time-evolution simulations, we calculate the expectation value of the order parameter (i.e., the net magnetization) $\mathcal{M}^z(t) = \lim_{L\rightarrow\infty} \sum_j \mel{\psi(t)}{\hat{\sigma}_j^z}{\psi(t)}/L$, which is related to the spontaneous breaking of the global $\mathbb{Z}_2$ symmetry hosted by the LR-TFIC, as well as the return rates with the two degenerate ground states at \(h = 0\), defined by
\begin{align}\label{eq:lambda1}
    \lambda^\pm_1(t) = - \lim_{L\rightarrow\infty} \frac{1}{L} \ln \left\lvert \braket{\psi^\pm}{\psi(t)} \right\rvert^2.
\end{align}
We further define the \textit{total} return rate as the minimum of these two values \cite{Jurcevic2017}: $r(t) = \min \left\lbrace \lambda^+_1(t), \lambda^-_1(t) \right\rbrace$. These return rates have a well-defined value in the thermodynamic limit, and can be calculated from the eigenvalues of the mixed iMPS transfer matrix \(\mathcal{T}^\pm(t)\) \cite{Zauner2015}, which we write in order of decreasing magnitude: \(\epsilon_n^\pm(t)\), \(n = 1, 2, \ldots\).
We define $\lambda_n^\pm(t) = -\ln \left| \epsilon_n^\pm(t) \right|$. For \(n = 1\), this expression is equivalent to the original definition of the return rates in Eq.~\eqref{eq:lambda1}~\cite{McCulloch:2008}.
DQPTs are defined as nonanalyticities in the total return rate \(r(t)\)~\cite{Heyl2013}, which generically correspond to \textit{manifold} crossings between \(\lambda^+_1(t)\) and \(\lambda^-_1(t)\) or \textit{branch} crossings between \(\lambda^\pm_1(t)\) and \(\lambda^\pm_2(t)\) \cite{VanDamme2022dynamical}.

An equivalent method of calculating the return rates~\eqref{eq:lambda1} involves calculating time-evolution simulations for each initial state \(\ket{\psi^\pm(t)}\), and evaluating $\lambda^\pm_1(t) = - \lim_{L\rightarrow\infty} L^{-1}\ln \left\lvert \braket{\psi^\pm(-t/2)}{\psi^+(t/2)} \right\rvert^2$, where we can obtain \(\ket{\psi^\pm(-t/2)}\) by taking the complex conjugate of \(\ket{\psi^\pm(t/2)}\), assuming the initial state is time-reversal symmetric.
By using this \textit{doubling trick}, if we perform a time-evolution simulation up to time \(T\) for each initial state, we can calculate the return rates up to time \(2T\) (but we can only calculate observables up to time \(T\)).

The dynamical critical point $h_\text{c}^\text{d}$ is defined such that if the quenched final value of the transverse-field strength $h_\text{f}<h_\text{c}^\text{d}$, the infinite-time average of the order parameter is nonzero, while if $h_\text{f}>h_\text{c}^\text{d}$, then the infinite-time average of the order parameter is zero \cite{Sciolla2010,Sciolla2011}. The value of $h_\text{c}^\text{d}$ also depends on the initial value $h_\text{i}$ of the quench parameter. For inverse-square interactions \(\alpha = 2\), the dynamical critical point of the LR-TFIC when $h_\text{i}=0$ is \(h_\text{c}^\text{d} \approx 1.85J\)~\cite{Halimeh2017dynamical,Zauner-Stauber2017probing}.
When we quench to a magnetic field below the dynamical critical point \(h_\text{f} < h_\text{c}^\text{d}\), we only observe DQPTs corresponding to branch crossings, and the order parameter does not pass zero, as shown in Fig.~\ref{fig:tfim}(a) for \(h_\text{f} = 1.25J\). Indeed, the order parameter remains significantly larger than zero throughout all accessible evolution times, indicating constrained dynamics.
However, if we quench above the dynamical critical point \(h_\text{f} > h_\text{c}^\text{d}\), we observe DQPTs corresponding to manifold crossings, each of which is associated with the order parameter crossing zero, as shown in Fig.~\ref{fig:tfim}(b) for \(h_\text{f} = 2.5J\).
This coincides to a crossover in the behavior of the quasiparticle excitations of this model~\cite{Vanderstraeten:2018,Halimeh2020quasiparticle}: below the equilibrium critical point (\(h_\text{c}^\text{e} \approx 2.5J\) in this case~\cite{Halimeh2017dynamical}), for \(h > h_\text{c}^\text{d}\), the lowest energy excitations are noninteracting domain walls, but for \(h < h_\text{c}^\text{d}\), a bound state of two domain walls has a lower energy than two vastly separated domain walls. This suggests that the transition between the dominance of branch DQPTs to their manifold counterparts is related to the confinement--deconfinement transition of the domain walls from bound states to freely propagating excitations.

\textbf{\textit{Two-dimensional quantum Ising model.---}}We now turn to the two-dimensional ($2$D) transverse-field Ising model on a square lattice, whose Hamiltonian is
\begin{align}\label{eq:2d-ising}
    \hat{H}_\text{2D-TFIM} = -J \sum_{\langle \mathbf{i},\mathbf{j}\rangle} \hat{\sigma}^z_\mathbf{i} \hat{\sigma}^z_\mathbf{j} - h \sum_\mathbf{i} \hat{\sigma}^x_\mathbf{i},
\end{align}
where $\mathbf{i}$ is a vector-index marking the location of a site on the square lattice. Unlike its one-dimensional counterpart, the $2$D Ising model with nearest-neighbor interactions has a finite-temperature phase transition at a critical (Onsager) temperature of $T_\text{c}=2J/\ln(1+\sqrt{2})$ \cite{Onsager1944}. On the square lattice, it has a quantum phase transition at the equilibrium critical point $h_\text{c}^\text{e}\approx3.004J$ \cite{deJongh1998,Bloete2002}.

\begin{figure}[t]
    \centering
    \includegraphics{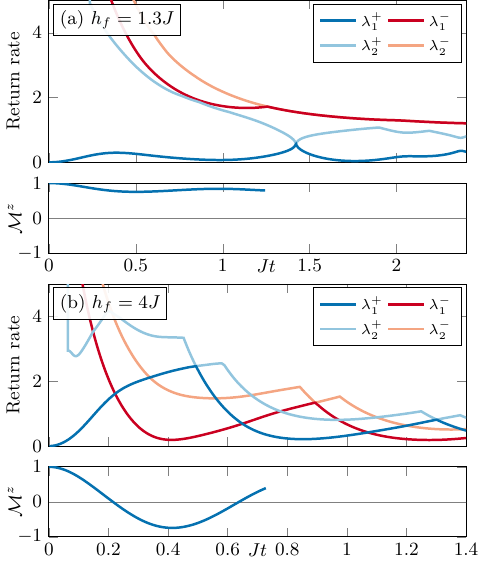}
    \caption{The return rates and order parameter of the quench of the transverse field in the two-dimensional square-lattice transverse-field Ising  model~\eqref{eq:2d-ising} on a cylindrical geometry with circumference \(L_y = 6\), from \(h_\text{i} = 0\) to (a)~\(h_\text{f} = 1.3J < h_\text{c}^\text{d}\) and (b)~\(h_\text{f} = 4J > h_\text{c}^\text{d}\).}
    \label{fig:2d-ising}
\end{figure}

As is standard for studying $2$D systems using MPS techniques, we put the system onto a cylindrical geometry, with a periodic circumference of finite width $L_y$, while the system is in the thermodynamic limit in the axial direction \(L_x = \infty\). For the results of this work, we set $L_y=6$ sites. Due to the finite length in the $y$-direction, finite-size fluctuations reduce the equilibrium quantum critical point to $h_\text{c}^\text{e}\approx2.93J$, and we determine, based on the order-parameter dynamics, that $h_\text{c}^\text{d} \approx 2.0J$ \cite{Hashizume2022dynamical}.
Our results are shown in Fig.~\ref{fig:2d-ising}: when we quench below the dynamical critical point (\(h_\text{f} = 1.3J < h_\text{c}^\text{d}\)), we only observe branch DQPTs directly and an order-parameter significantly above zero indicating constrained dynamics, while when we quench above the dynamical critical point (\(h_\text{f} = 4J > h_\text{c}^\text{d}\)), we only observe manifold DQPTs corresponding to order parameter zeros.

As such, the same picture holds as in the case of the LR-TFIC: in the presence of confinement, branch DQPTs arise and dominate the return rate concurrently with the order parameter exhibiting constrained dynamics and never passing zero, while in the deconfined phase, manifold DQPTs dominate the return rate and correspond directly to order-parameter zeros.

\textbf{\textit{Spin-$S$ $\mathrm{U}(1)$ quantum link model.---}}Let us now consider a paradigmatic gauge theory, the spin-\(S\) \(\mathrm{U}(1)\) quantum link model (QLM) with Hamiltonian \cite{Chandrasekharan1997,Wiese_review,Kasper2017}
\begin{align}\nonumber
    \hat{H}_\text{QLM} =&-\frac{\kappa}{2} \sum_i \left( \hat{\phi}_i^\dagger \hat{s}^+_{i,i+1} \hat{\phi}_{i+1} + \text{H.c.} \right) \\\label{eq:qlm}
    &+ \mu \sum_i (-1)^i\hat{\phi}^\dagger_i \hat{\phi}_i + \frac{g^2}{2} \sum_i \big(\hat{s}^z_{i,i+1}\big)^2,
\end{align}
where \(\hat{\phi}^\dagger_i\) (\(\hat{\phi}_i\)) is a fermionic creation (annihilation) operator acting on matter site \(i\), even (odd) sites represent (anti)particles (for antiparticles, the role of creation and annihilation is swapped), \(\hat{s}^\pm_{i,i+1}\) and \(\hat{s}^z_{i,j+1}\) are spin-\(S\) operators acting on the gauge link between matter sites \(i\) and \(i+1\), representing the gauge and electric-field operators, respectively, \(\kappa\) is the hopping energy, \(\mu\) is the fermion mass, and \(g\) is the gauge-coupling strength, which controls the confinement of the particles.

\begin{figure}[t]
    \centering
    \includegraphics{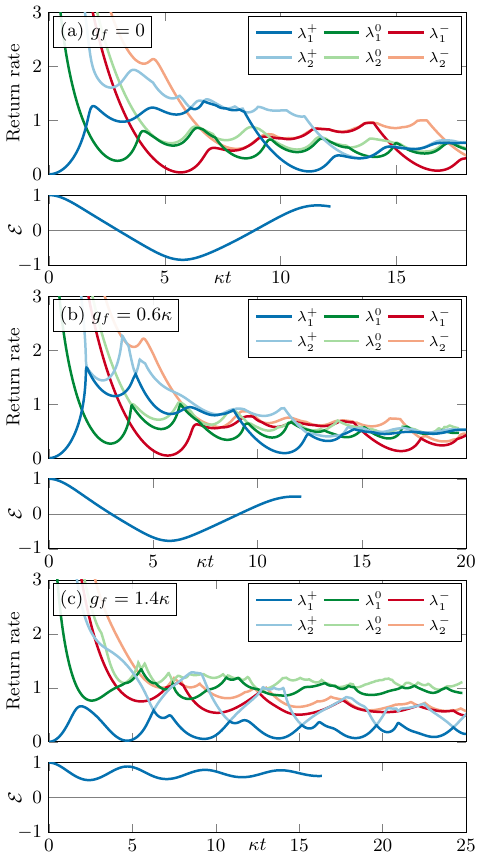}
    \caption{The return rates and order parameter of the quench of the spin-$1$ \(\mathrm{U}(1)\) QLM~\eqref{eq:qlm}, \(\mu = 0.1\kappa\), from the positive extreme vacuum \(\ket{\psi^+}\) to (a)~\(g_\text{f} = 0\), (b)~\(g_\text{f} = 0.6\kappa\) and (c)~\(g_\text{f} = 1.4\kappa\).}
    \label{fig:qlm}
\end{figure}

The generators of the \(\mathrm{U}(1)\) gauge symmetry are
\begin{equation}
    \hat{G}_i =\hat{s}^z_{i,i+1}- \hat{s}^z_{i-1,i}-\hat{\phi}^\dagger_i \hat{\phi}_i - \frac{(-1)^i-1}{2},
\end{equation}
which are a discrete analog of Gauss’s law.
We work in the physical sector where \(\hat{G}_i \ket{\psi} = 0\), \(\forall i\).

In this work, we consider \(S = 1\) --- the effect of the gauge coupling \(g\) is trivial for \(S = 1/2\) as then $(\hat{s}^z_{i,i+1})^2=\mathds{1}/4$ is an inconsequential constant energy term --- where there are three different vacua in the physical sector, which can be written in a basis characterized by the \(z\) projection of the spin \(m_z\) of each link flux: $\hat{s}^z_{i,i+1} \ket{\psi^{m_z}} = m_z \ket{\psi^{m_z}},\,\forall i$, where \(m_z = -1, 0, +1\) (we will write \(m_z = \pm\) as an abbreviation of \(\pm 1\)).
We then numerically simulate a sudden quench from the ``extreme'' vacuum \(\ket{\psi^+}\) to a finite \(g\) and \(\mu\) using the same numerical methods as in the previous sections with a time-step of \(0.01/\kappa\), and we calculate the return rates $\lambda_1^{0,\pm}(t)$ for the three vacua, with the total return rate $r(t)=\min \left\lbrace \lambda^+_1(t), \lambda^0_1(t), \lambda^-_1(t) \right\rbrace$ being their minimum, as well as the electric flux $\mathcal{E}(t) = \lim_{L\rightarrow\infty}\sum_i \mel{\psi(t)}{\hat{s}^z_{i,i+1}}{\psi(t)}/L$, which is an order parameter associated with a global $\mathbb{Z}_2$ symmetry related to charge-parity conservation \cite{Coleman1976,Rico2014}.

The gauge coupling $g$ tunes the strength of confinement. For small \(g\), confinement is weak: this is shown in the behavior of the flux in our quenches to \(g_\text{f} = 0\) and \(0.6\kappa\) in Fig.~\ref{fig:qlm}(a,b).
In these quenches, the flux oscillates close to the extreme values \(\pm 1\): this is indicative of the free propagation of single deconfined matter particles.
We also observe primarily manifold DQPTs, with a zero crossing of the flux corresponding to a maximal overlap of the wave function with the middle vacuum. At larger values of \(g\), such as in the quench to \(g_\text{f} = 1.4\kappa\) in Fig.~\ref{fig:qlm}(c), confinement is strong, as indicated by the flux oscillating close to its initial value. In this case, we only observe branch DQPTs, and no manifold DQPTs at all. Once again we see how the dominance of branch DQPTs in the return rate coincides with strong confinement.

It is worth mentioning that for the quenches in the spin-$1$ \(\mathrm{U}(1)\) QLM~\eqref{eq:qlm}, we have to calculate time-evolution simulations for quenches of all three initial states \(\ket{\psi^{0,\pm}(t)}\).
However, since the quench of the middle vacuum \(\ket{\psi^0}\) is more numerically expensive than those of the two extreme vacua \(\ket{\psi^\pm}\), we cannot reach as long simulation times for the former as in the latter.
Hence, the total time for which we can calculate all of the return rates in Fig.~\ref{fig:qlm} is slightly less than double the final time of the order parameter.

\textbf{\textit{Discussion and outlook.---}}We have shown how the prevalence of \textit{branch} DQPTs in the return rate, which involve nonanalyticities in the overlap to the initial state, is an indicator of confinement. We have demonstrated this using iMPS calculations involving the power-law interacting quantum Ising chain, quantum Ising model on a square lattice, and spin-$S$ $\mathrm{U}(1)$ QLMs of lattice QED. Our numerical simulations show a clear picture: whereas in the deconfined phase mostly manifold DQPTs arise that have a direct connection to order-parameter zeros, under strong confinement only branch DQPTs arise, coinciding with an order parameter exhibiting constrained dynamics near its initial value.

Given the high level of control and precision in today's quantum simulators, our conclusions can be tested in ion-trap and cold-atom setups~\cite{Jurcevic2017,Su2022}. Whereas in Ref.~\cite{Jurcevic2017} only those regimes in which manifold DQPTs occur were probed, quenches within the ferromagnetic phase of the long-range quantum Ising chain would be able to exhibit branch DQPTs while simultaneously observing constrained order-parameter dynamics, the latter as explained in Ref.~\cite{Liu2019}. For the spin-$1$ $\mathrm{U}(1)$ QLM, recent experimental proposals have shown how this can be realized using bosonic Dysprosium atoms in optical superlattices, which would facilitate the verification of our conclusions in lattice gauge theories \cite{osborne2023spins}.

It would be interesting to understand how confinement gives rise to branch DQPTs. Given that the models considered here are nonintegrable and their far-from-equilibrium dynamics is highly nonperturbative, analytic insights are limited, and numerical methods become our best approach to understanding the physics. Nevertheless, one can speculate that confinement gives rise to fundamentally \textit{far-from-equilibrium} criticality in the form of branch DQPTs, distinct from the rather equilibrium-like criticality of manifold DQPTs, which are directly associated to the order-parameter dynamics changing sign. Furthermore, and unsurprisingly, it seems that the criticality of branch DQPTs involves much higher-order excitations than the low-lying quasiparticles due to confinement such as bound domain walls in the LR-TFIC, because low-energy approximations of the LR-TFIC such as the two-kink model \cite{Liu2019} do not capture branch DQPTs.

It is worth mentioning that we have intentionally chosen models with unbroken global symmetries, in order to be able to properly define and distinguish between manifold and branch DQPTs. Of course, there are models such as the Ising chain with both transverse and longitudinal fields and the spin-$1/2$ $\mathrm{U}(1)$ QLM with a topological $\theta$-term in which the global symmetry is explicitly broken. In such a case, the initial-state manifold is nondegenerate, and the return rate involves projecting on a single initial state, thereby preventing a distinction between manifold and branch DQPTs. This is something we have intentionally avoided for clarity. With that said, classifying confinement in such models through DQPTs would also be interesting, and is left for future work.

\bigskip

\begin{acknowledgments}
\textbf{\textit{Acknowledgments.---}}J.C.H.~acknowledges funding within the QuantERA II Programme that has received funding from the European Union’s Horizon 2020 research and innovation programme under Grant Agreement No 101017733, support by the QuantERA grant DYNAMITE, by the Deutsche Forschungsgemeinschaft (DFG, German Research Foundation) under project number 499183856, funding by the Deutsche Forschungsgemeinschaft (DFG, German Research Foundation) under Germany's Excellence Strategy -- EXC-2111 -- 390814868, and funding from the European Research Council (ERC) under the European Union’s Horizon 2020 research and innovation programm (Grant Agreement no 948141) — ERC Starting Grant SimUcQuam. I.P.M.~acknowledges funding from the National Science and Technology Council (NSTC) Grant No.~122-2811-M-007-044. Numerical simulations were performed on The University of Queensland’s School of Mathematics and Physics Core Computing Facility \texttt{getafix}.
\end{acknowledgments}

\bibliography{biblio}

\end{document}